\documentclass[%
 reprint,
 amsmath,amssymb,
 aps,
 prb
]{revtex4-2}

\usepackage{graphicx}
\usepackage{dcolumn}
\usepackage{bm}
\usepackage{color}
\usepackage[colorlinks=true,linkcolor=blue,allcolors=blue]{hyperref}

\begin{document}

\preprint{prb}

\title{Electron transport in disordered insulating lattice under nonlinear electric field}

\author{Kunal Mozumdar}
 \author{Herbert F. Fotso}%
\affiliation{%
 Department of Physics, University at Buffalo, USA}
 
\author{Jong E. Han}%
 \email{jonghan@buffalo.edu}
\affiliation{%
 Department of Physics, University at Buffalo, USA}




\date{\today}

\begin{abstract}
Transport in disordered systems often occurs via the variable range hopping (VRH) in the dilute carrier density limit, where electrons hop between randomly distributed localized levels. We study the nonequilibrium transport by a uniform DC electric field on a one-dimensional insulating tight-binding chain with the on-site disorder, using a disordered-lattice calculation and the coherent potential approximation. We develop a theory of electric-field-assisted variable range hopping as a mechanism for nonlinear transport in a disordered chain. Our disordered-lattice calculations of the electron propagation distance and the electron mobility determine the range of the variable range hopping as $\Delta < W \lesssim 2\Delta$ in the gap $\Delta$. We further propose a nonlinear scaling of the conductivity by an electric field by extending Mott's variable range hopping. The nonlinear conductivity of an electronic lattice model follows the scaling law $\sigma(E) \propto \exp[-(E_0/E)^{\nu}]$ with the exponent $\nu = 1/3$ in one dimension for the VRH. We also discuss the experimental relevance of temperature-dependent nonlinear current-voltage relation.
\end{abstract}

\maketitle


\section{\label{sec:level1}Introduction}
The problem of disorder in solid-state systems has long been a significant area of focus in condensed matter physics. Anderson's pioneering work \cite{anderson1958absence} demonstrated that in a lattice with disordered potentials, electrons become localized in certain regions of the lattice, known as the Anderson localization. This phenomenon has been studied extensively~\cite{RevModPhys.57.287,thouless1974electrons,mott1967electrons,Mott01041969,shklovskii2013electronic,anderson201050,cutler1969observation} in the context of electronic systems in equilibrium, and the concept has since been expanded to encompass various wave phenomena \cite{sheng1990scattering, kirkpatrick1985localization, john1984electromagnetic,lagendijk2009fifty}. 

Neville Mott, a decade after Anderson's work, \cite{mott1967electrons, mott1971conduction, mott1975anderson} discovered an electronic transport mechanism in a regime of low carrier density where thermal excitations are not sufficient for electrons to reach the conduction band. He proposed that the electrons hop to localized levels at varying distances but between sites closer in energy. Mott derived ~\cite{mott1968conduction} the following expression for the conductance
\begin{equation} \label{mott eq}
    \sigma = \sigma_0 \exp[-(T_0/T)^{1/(d+1)}]
\end{equation}
with $d$ the dimensionality of the system, by optimizing the Miller-Abraham's formula for the hopping rate ${\cal W}$ at temperature $T$ between levels separated by distance $R$ and energy excitation by $\Delta\epsilon$,
\begin{equation}
{\cal W}(R,T)\propto \exp\left[-\frac{2R}{\xi}-\frac{\Delta\epsilon}{k_BT}\right],    \label{eq:MillerAbraham}
\end{equation}
where $\xi$ is the localization length, $k_B$ the Boltzmann constant. For a one-dimensional system, Mott's scaling Eq.~(\ref{mott eq}) has the exponent of $1/2$ with $T_0 = 2[k_{\text{B}}\rho(\epsilon_{\text{F}})\xi]^{-1}$ where $\rho(\epsilon_{\text{F}})$ is the DOS at the Fermi level $\epsilon_F$. This phenomenon, known as the variable-range-hopping (VRH), or simply as the hopping mechanism, has been observed experimentally in various materials \cite{doi:10.1021/nn101376u,evans2017mono,PAASCH200297,Apsley1974-APSTFO,Rahman_2010} and theoretically studied extensively using resistor networks based on the semi-classical percolation problem~\cite{10.1063/10.0034343,lee1984variable,shklovskii2013electronic,PhysRevB.69.035413}. 

The purpose of this work is to deepen the theoretical understanding of the VRH in the nonlinear transport regime in an electronic lattice where the electron dynamics is explicitly considered. While the role of strong fields in condensed matter is receiving a great deal of interest lately in nanoscale device applications, the theoretical framework has only been developed in recent decades, particularly for quantum transport far from equilibrium. The problem is many-fold: theoretically, anything outside equilibrium cannot rely on textbook statistical mechanical principles such as Gibbs' ensemble, leaving the theory only with quite formidable options such as the Keldysh diagrammatic theory in such a regime \cite{aoki2014nonequilibrium,haug2008quantum}. From the phenomenological perspective, nonequilibrium has quite a distinct reference state from an equilibrium that has entangled quantum and classical excitations that are often hard to distinguish. In the past, \textit{ad hoc} scaling relations such as those obtained by replacing thermal energy with field-driven energy in thermodynamics relations~\cite{PhysRevB.70.235120} have often been used. In this work, we instead treat the disordered nonequilibrium limit rigorously by considering the electronic lattice model by solving the Keldysh theory in a straightforward manner with the disorder averaging coming only at the end of the calculation. Therefore, we make a definitive statement for the scaling behavior of high-field conductivity in one of the most basic nonequilibrium models that contains the following minimum essential ingredients: lattice, Fermi-statistics, electric field, disorder, and dissipation. We test some notable results \cite{Mott01071970,shklovskii1973,rodin2009numerical, prigodin1980one, PhysRevLett.95.166604,zaccone2016,SENOUCI199623} that discuss the transport of electrons in disordered media in an arbitrary electric field.

In our previous work \cite{mozumdar2025}, we addressed the spectral properties of a field-driven tight-binding model and validated Mott's scaling of the VRH through Keldysh Green's function method. The calculations highlighted that the electric field results in delocalization from the Anderson localization to a crossover to the Wannier Stark effect at a very strong electric field, by using the spectral properties of the system. We also observed that in equilibrium, the disordered-lattice calculations showed exponential Lifshitz tails, which are further broadened by the effect of the electric field, and it was argued that this is crucial for understanding the transport behavior of disordered chains. Furthermore, lattice-explicit calculations have revealed that the nonequilibrium distributions in the steady-state show highly nontrivial non-thermal evolution.

In the present paper, we focus on the electronic transport of a disordered strongly-driven electron lattice. We present a simple Anderson model of disorder consisting of a tight-binding chain and an on-site disordered potential and apply a DC electric field. We couple the chain to an infinite fermionic reservoir at each site, which ensures a steady-state solution through dissipation. We compute the nonequilibrium Green's functions in the steady-state limit and compute the transport quantities. We explicitly demonstrate how the hopping mechanism evolves with an electric field in and out of the VRH limit by studying the propagation length. In nonequilibrium, we derive Mott's scaling law as a function of the electric field, which we corroborate with Keldysh Green's function theory. We also investigate the $IV$ (current-voltage) evolution in the VRH regime for nontrivial crossover from the hopping to Ohmic regime, which can be readily compared for nonlinear transport in quasi-one-dimensional systems. We finally discuss experimentally relevant temperature dependence of the field-activation of nonlinear conductance.  

\begin{figure}
\includegraphics[width=\linewidth]{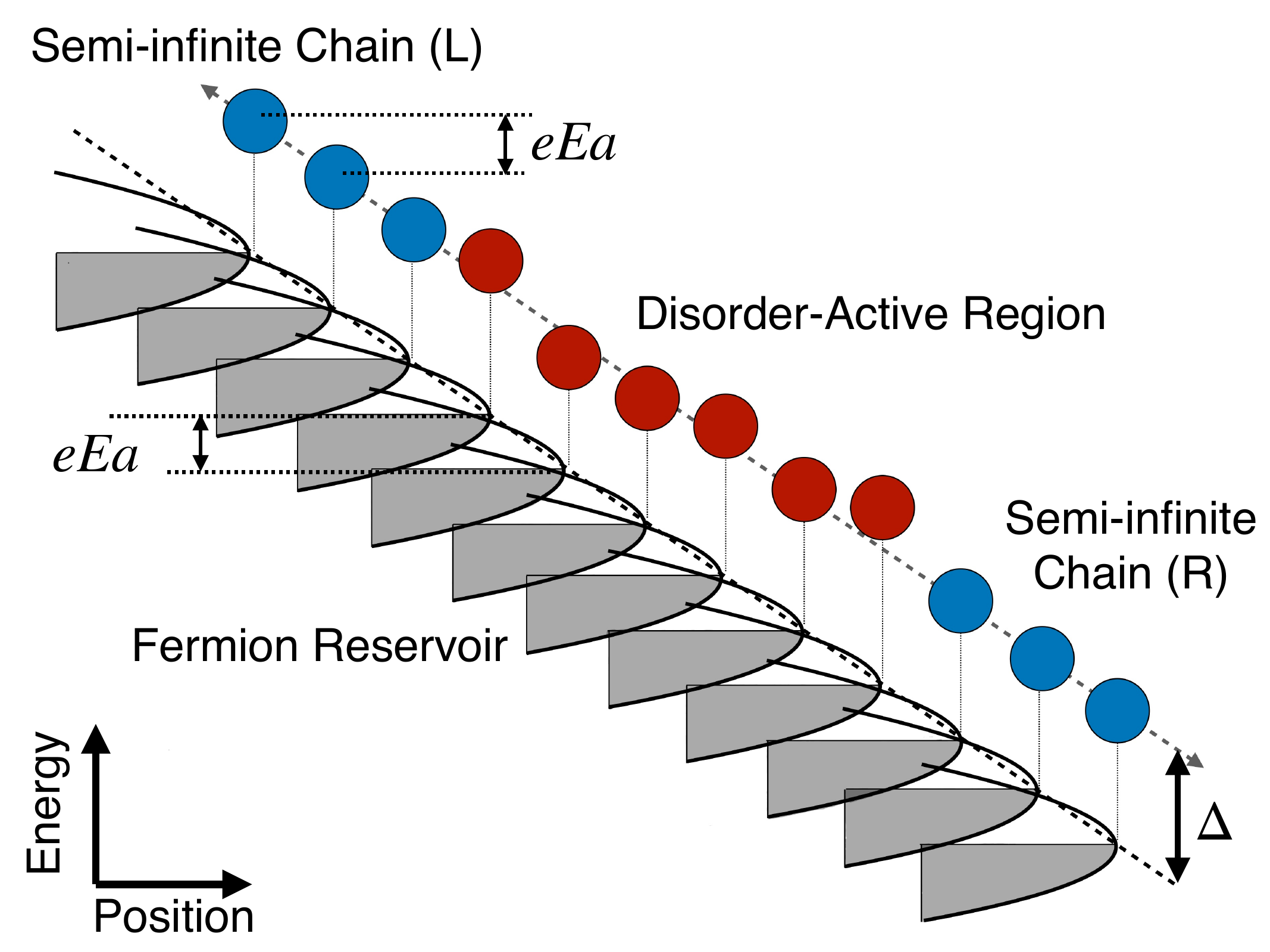}
\caption{Infinite tight-binding chain set-up with the electric field. The disorder-active region (red circles) of the chain consists of $N=501$ sites with Anderson disorder in the range $(-W,W)$. On both ends of the active region, we have a semi-infinite chain without disorder that acts as leads (blue circles). Every site is coupled to a fermionic chain reservoir, which helps maintain the steady-state limit out of equilibrium. Electrostatic potential $-lEa$ is applied across the entire chain, including the leads and the reservoir levels. The main chain is kept above the Fermi level, marked by the shaded region by a gap $\Delta$.}
\label{fig:1}
\end{figure}  

The rest of the paper is summarized as follows. In section-\ref{sec:level2}, we summarize our quantum mechanical model introduced in our previous study~\cite{mozumdar2025} and discuss our calculations of Green's functions, also discussed in great detail in the first paper, as well as transport properties. In section-\ref{sec:level3}, we discuss our results for the disordered lattice case in equilibrium and nonequilibrium. We highlight some key aspects of our calculation against the approach of coherent potential approximation (CPA) for the disordered potential. Finally, in section-\ref{sec:level4}, we discuss our results and conclude. 

\section{\label{sec:level2}Disordered Chain under DC bias}

Our model~\cite{mozumdar2025} consists of an infinite tight-binding chain, with a constant electric field applied to all sites and the central region (red circles) as the disorder-active region, as shown in Fig.~\ref{fig:1}. Each site at position $l$ in the main chain is at an electrostatic potential $-lEa$, $a$ being the lattice constant (set to unit length). With the setup, physical (gauge-covariant) observables such as electron density and current are translationally invariant in the zero disorder limit without any scattering between the disorder-active and the lead regions. Each site of the chain is coupled to fermion chains which act as a reservoir~\cite{hanPRB2013} as shown in Fig.~\ref{fig:1}. The Hamiltonian hence can be written as 
\begin{eqnarray}
\mathcal{H} = && -t\sum_{l={-\infty}}^{\infty} (d_l^{\dagger}d_{l+1} + d_{l+1}^{\dagger}d_{l})+ \sum_l \epsilon_l d_l^{\dagger}d_{l} \nonumber\\
&& 
+ \sum_{l\alpha} \epsilon_{l\alpha} c_{l\alpha}^{\dagger}c_{{l\alpha}} - \frac{g}{\sqrt{L}}\sum_{l\alpha} (c_{l\alpha}^{\dagger}d_{l} + h.c.)
\label{eq:1}
\end{eqnarray}
The first two terms of the Hamiltonian denote the main chain with electron creation (annihilation) operator $d^{\dagger}_l$ $(d_l)$ which is coupled to the fermion reservoir that consists of fermion states with creation (annihilation) operator $c^{\dagger}_{l\alpha}$ $(c_{l\alpha})$, with $\alpha$ being the continuum band index, described by the last two terms. $\epsilon_l$ is the level energy for the orbital at the $l$-th site on the main chain, while $\epsilon_{l\alpha}$ is the continuum energy of the bath states attached to the $l$-th site with the continuum index $\alpha$. The reservoir acts as a sink for excess energy generated by the electric field in the main chain. 

The tight-binding parameter, $t$ in our system (set to $1$ unit of energy) and the site energy $\epsilon_l$ is given as
\begin{equation}
    \epsilon_l = 2t + \Delta + V_l - l\cdot eEa
\label{eq:2}
\end{equation} 
where $E$ denotes the uniform DC electric field, $\Delta$ is the gap and $V_l$ is the on-site random disordered potential. The elecric charge $e$ and the lattice constant $a$ are set to 1, for the unit of charge and length, respectively. The bath spectrum $\epsilon_{l\alpha}$ is similarly given as $\epsilon_\alpha-lE$ with the identical spectra $\epsilon_\alpha$ for every site $l$. Our system is considered to be a discretized limit of an insulator with the dispersion relation $\epsilon_p = p^2/2m + \Delta$ (with $1/(2m)=t=1$), that is displaced from the Fermi level as shown in Fig.~\ref{fig:1}. We set the gap $\Delta = 0.3$ throughout our calculations. 

The disorder potential is set as 
\begin{equation}
    V_l=\left\{\begin{array}{ll}
    \mbox{random in }[-W,W] & \mbox{for }-N/2\leq l\leq N/2 \\
    0 & \mbox{otherwise}
    \end{array}
    \right..
\end{equation}
The nature of the disorder is similar to Anderson disorder, drawn from a uniform random distribution $P(V_l) = (2W)^{-1}\Theta(W - \left|V_l\right|)$ with disorder strength $W$. This defines the disorder-active region in the main chain as shown in Fig. \ref{fig:1}. We chose $N =501$ active sites, allowing us to attain a bulk limit within the bounds of computational feasibility. 

The reservoir consists of a fermion chain of length $L$ with identical dispersion $\epsilon_{\alpha}$ for each site. Its density of states is assumed to be structureless, constant, and with an infinite bandwidth. Mixing of the main orbitals with the reservoir is given by the hybridization function $\Gamma(\omega) = \pi g^2 L^{-1}\sum_{\alpha} \delta(\omega - \epsilon_{\alpha}) \equiv \Gamma$ in the infinite bandwidth limit. This parameter $\Gamma$, also referred to as dissipation rate in the text, denotes the energy level broadening and dephasing rate of electrons. The reservoir is kept at a temperature $T$ with the Fermi level shifted by the electrostatic potential $-lE$. Hence the electron distribution at each site $l$ is given by the Fermi-Dirac function as $f_{\rm FD}(\omega-lE)=[1+e^{(\omega-lE)/T}]^{-1}$ in the reservoir. (We set the Boltzmann constant $k_B=1$.) We incorporate the dissipation effects of the reservoir exactly by using the Keldysh Green's function method and analyze its effect on the transport properties in the subsequent sections.  

\subsection{\label{sec:level2-1} Lattice Calculation of the conductance}
We numerically calculate the full retarded Green's function of the system for each disordered configuration. The Green's function of an infinite chain can be written in terms of the self-energy from the disorder-free regions. More detailed discussions on the method can be found in our previous work~\cite{mozumdar2025}. The general form of the $N \times N$ retarded Green's function matrix is given by
\begin{eqnarray}\label{eq:3}
    [\mathcal{G}^{R}(\omega)^{-1}]_{ij}  & = & (\omega - \epsilon_i + i\Gamma)\delta_{ij} + t\delta_{|i-j| = 1} \\
& & 
 - t^2F^R_{-}(\omega - NE/2)\delta_{i,-N/2}\delta_{j,-N/2} \nonumber \\
& &    
 - t^2F^R_{+}(\omega + NE/2)\delta_{i,N/2}\delta_{j,N/2}] \nonumber 
\end{eqnarray}
and the lesser Green's function is given by
\begin{equation}\label{eq:4}
\mathcal{G}^{<}_{ij}(\omega) = \sum_{k=1}^N \mathcal{G}^{R}_{ik}(\omega) \Sigma^{<}_k(\omega) [\mathcal{G}^{R}_{jk}(\omega)]^{*} 
\end{equation}

In Eq. (\ref{eq:3}), the term $-i\Gamma$ is the retarded reservoir self-energy causing broadening of levels and $t^2F^R_-$  and $t^2F^R_+$ are the self-energy terms of the semi-infinite leads attached on the left and right side of the disordered active sites, respectively, as depicted in Fig. \ref{fig:1}. The retarded Green's function of the right/left chains $F^R_\pm$, evaluated at the terminating orbital of the semi-infinite chain, are computed iteratively \cite{li2015electric} as
\begin{equation} \label{eq:Fret}
    [F^{R}_{\pm}(\omega)]^{-1} = \omega + i\Gamma - (2t + \Delta) - t^2F^{R}_{\pm}(\omega \pm E). 
\end{equation}
The Green's functions on the disorder-active sites are obtained by matrix inversion using sparse matrix routines. In Eq.~(\ref{eq:4}) the term $\Sigma^{<}_{k}(\omega) = 2i\Gamma f_{\rm FD}(\omega+kE) + t^2\delta_{k,-N/2}F^{<}_{-}(\omega - NE/2) + t^2\delta_{k,N/2}F^{<}_{+}(\omega+NE/2)$ denotes the lesser self-energy at site $k$. Similarly, the lesser Green's functions $F^<_\pm$ for the semi-infinite chains are computed iteratively as 
\begin{equation} \label{eq:Fless}
    F^{<}_{\pm}(\omega) = |F^{R}_{\pm}(\omega)|^2 [2i\Gamma f_{\rm FD}(\omega) + t^2F^{<}_{\pm}(\omega \pm E)] 
\end{equation}

Finally, physical observables such as the electron occupation and current are computed from the Green's functions. The local current $J_l$ at site $l$ \cite{li2015electric} is given by
\begin{equation}
J_l = -t \int [\mathcal{G}^{<}_{l,l+1}(\omega) - \mathcal{G}^{<}_{l+1,l}(\omega)]\frac{d\omega}{2\pi}
\label{eq:5}
\end{equation}
and the local occupation number by
\begin{equation}
n_l = \int \mathcal{G}^{<}_{ll}(\omega)\frac{d\omega}{2\pi i}.
\label{eq:6}
\end{equation}
While $n_l$ and $J_l$ are independent of $l$ at zero disorder, disorder introduces fluctuations in space due to the random potential on each site. We obtain all physical quantities as averaged over the disorder configurations and the disorder-active sites. We chose central $201$ sites to remove any possible scattering effect from the reservoir. This calculation is then repeated and averaged over $3000$ independent disorder configurations. \\
The following parameters are tuned - disorder strength $W$, electric field $E$, temperature $T$, and dissipation $\Gamma$ to study transport quantities such as the local conductivity defined as :
\begin{equation}
\sigma = \frac{\langle J_l\rangle}{E}
\end{equation}
and the mobility is defined as
\begin{equation}
    \mu = \frac{\sigma}{\langle n_l\rangle} 
\end{equation}
where for simplicity $\langle f \rangle = \sum_{i} [\int dv_i P(v_i) f_i]$ denote both averaging over spatial sites and disorder configurations. Similar to \cite{mozumdar2025}, we explore the electric field range in $E=10^{-4}$ to $0.1$ which corresponds to $10$ to $10^4 \text{ kV/cm}$ that includes typically experimental scales. The dissipation parameters are considered smaller than lattice parameters and we keep $\Gamma$ in the order of $\Gamma\sim 10^{-4}$.

In the studies of the electron transport in disordered media~\cite{shklovskii2013electronic}, the Miller-Abraham's relation, Eq.~(\ref{eq:MillerAbraham}), has been the cornerstone of statistical theories, including Mott's. Our approach puts the quantum mechanical problem by including the effects of electric field, disorder, and dissipation in a first-principles manner and solves the Schr\"odinger equation of an electronic lattice directly by the Keldysh Green's function method. More specifically, the first term in the argument of Eq.~(\ref{eq:MillerAbraham}) for the wavefunction overlap is accounted for by the retarded Green's functions $|{\cal G}^R_{ij}(\omega)|^2$ in Eq.~(\ref{eq:4}), and the thermal factor in Eq.~(\ref{eq:MillerAbraham}) is represented by the lesser quantities ${\cal G}^<_{ij}(\omega)$ or $\Sigma^<_k(\omega)$ in the framework of the Keldysh theory.

As a benchmark, we complement the full-lattice calculation with the coherent potential approximation (CPA) \cite{elliott1974theory,janivs2021dynamical,dohner2022nonequilibrium} approach. This method is similar to the dynamical mean field theory \cite{georges1996dynamical,aoki2014nonequilibrium,10.3389/fphy.2020.00324} (DMFT) where the spatial inhomogeneity of the disordered lattice is replaced by a complex self-energy which represents the local effective medium potential. This is evaluated self-consistently by embedding an impurity with a random potential drawn in the range of $-W \leq V \leq W$ and then computing the average Green's function, which is mapped to the effective medium Green's function. We employ the nonequilibrium steady-state calculation similar to the one discussed in the earlier works of one of the authors \cite{li2015electric}, along with the CPA approach to incorporate the effects of disorder to compare the disordered-lattice calculations. The CPA method is described in detail in \cite{mozumdar2025}. The transport variables like the charge occupation and the conductivity are computed in the same manner as the disordered-lattice calculations, although they are only defined locally at the impurity site.

\begin{figure*}
    \centering
    \includegraphics[width=\linewidth]{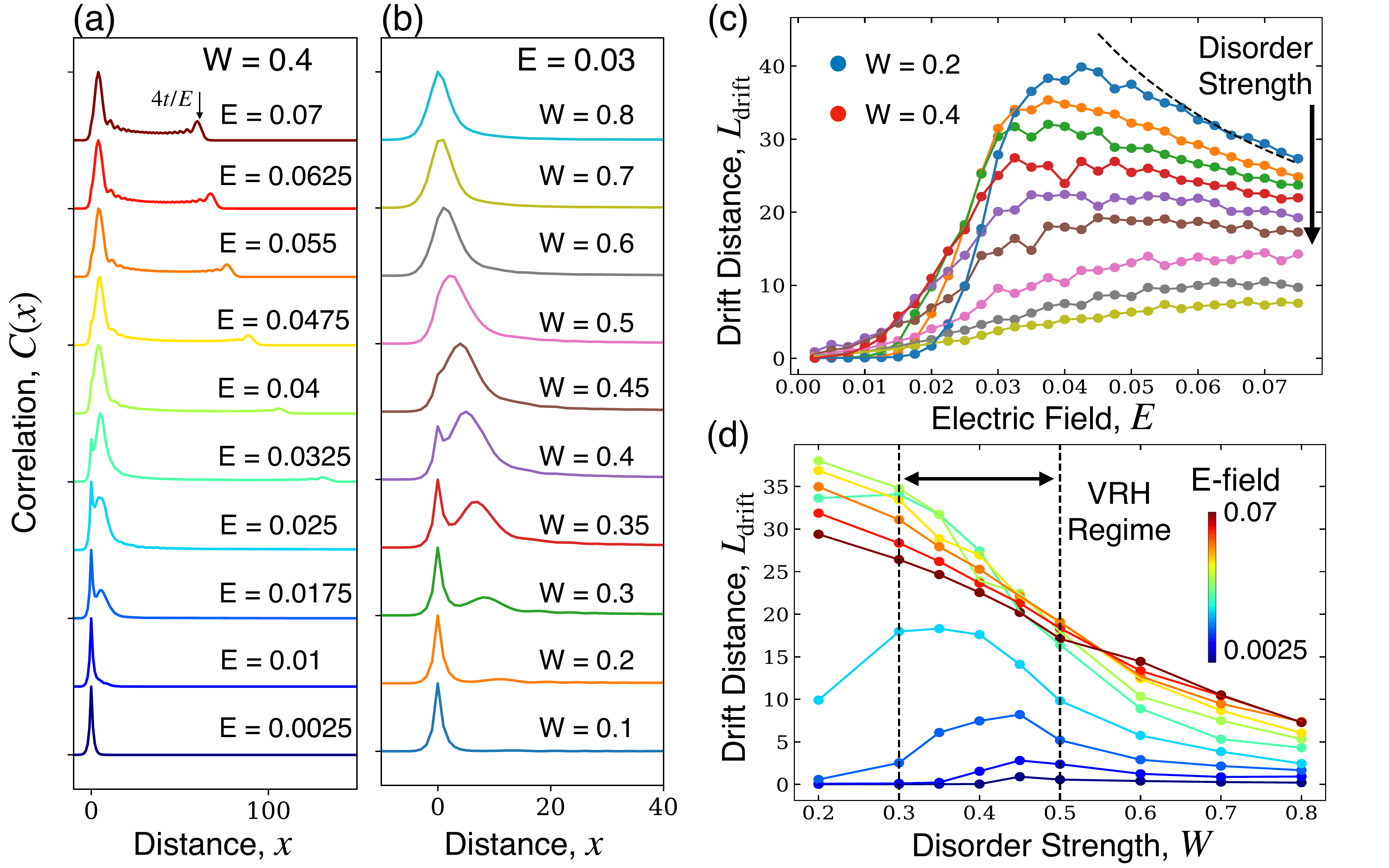}
    \caption{(a) Spatial correlation function $C(x)$ plotted for different electric fields at disorder strength $W = 0.4$, for $\Gamma = 0.0005$. As the electric field $E$ increases, $C(x)$ widens, indicating the tendency of delocalization. As $E$ increases further, the signature of Bloch oscillations appears from large $x$. The range of Bloch oscillations scales well with $4t/E$ (as marked by the arrow). (b) Spatial correlation function at intermediate electric field ($E = 0.03$) for different disorder values. The width of $C(x)$ reaches a maximum in the VRH regime. (c) Drift distance $L_{\rm drift}$ vs against electric field for disorder strengths 
    [$W \in \{0.2, 0.3, 0.35, 0.4, 0.45, 0.5, 0.6, 0.7, 0.8\}$ in the order indicated by the arrow]. $L_{\rm drift}$ initially increases with $E$ due to the field-activation~\cite{FrenkelPR1938}, and then decreases due to Bloch oscillations $x\sim 4t/E$ (marked by the dashed line). The initial rise at small fields is the strongest in the VRH regime. (d) Drift distance $L_{\rm drift}$ vs disorder strength at various electric fields ranging from $E=0.0025$ (dark blue) to $E = 0.07$ (dark red) with a spacing of $\Delta E = 0.0075$. The VRH range is demarcated by the black dashed lines.
}
    \label{fig:2}
\end{figure*}

\section{\label{sec:level3}Results}
\subsection{\label{sec:level3-1}Electron Drift in Disordered Lattice}

In this subsection, we address the statistical properties in the hopping transport by computing the correlation function $C(x)$ and the average electron drift distance $L_{\rm drift}$. This correlation analysis elucidates the statistical overlap between wave functions and gives a direct insight into the hopping behavior such as the conductance and mobility.

The wave-function overlap of an electron that propagates over the distance $x$ can be computed via the correlation function $C(x)$
\begin{equation}
    C(x) = \left\langle \frac{1}{M} \sum_{m=0}^{M} |\mathcal{G}^{R}_{m+x,m}(\omega = -mE)|^2 \right\rangle
\end{equation}
We compute $C(x)$ in the disordered-lattice calculation numerically by computing the disorder-averaged correlation of the non-diagonal matrix elements terms of the retarded Green's function at the local Fermi energy $-mE$ averaged over each $m$-th site, since the transport is carried over states close to the Fermi energy. Here, $x$ denotes the spatial index ($0 \leq x \leq 150$) across the 1D lattice and $M=201$ denotes the number of sites chosen for spatial averaging around the center of the lattice of $N=501$ sites. This quantity $C(x)$ is averaged over multiple disorder configurations ($\sim 2000$ different configurations). 

For strongly localized states, this quantity decays exponentially and for a delocalized or extended state, it spreads over the entire lattice.  Fig. \ref{fig:2}(a) depicts the behavior of $C(x)$ for increasing electric field at the disorder strength $W=0.4$ inside the VRH regime. At low electric fields, the wave functions are typically localized; but at higher fields, the wave function shows evidence for the Bloch oscillations. Although the Bloch oscillation localizes the electron motion via the Brillouin zone averaging, it is only possible when the electron traverses the Brillouin zone. Therefore, the emergence of the Bloch oscillation can be taken as a sign of delocalization. The spread of $C(x)$ behaves inversely proportional to the electric field and at high fields $E \gtrsim 0.4$, $C(x)$ extends to the distance $x_B=4t/E$ (as marked by an arrow for $E=0.07$). The width of the Bloch oscillation can be easily understood as the travel distance of an electron driven by an electric field as
\begin{equation}
x_B=\int_{v_g>0}v_g(k)\frac{dk}{\dot{k}}=\frac{4t}{E},
\end{equation}
with the group velocity $v_g(k)=2t\sin(k)$ and the semi-classical equation of motion $\hbar\dot{k}=eEa$. This behavior corroborates with the discussion~\cite{mozumdar2025} that, at high fields, the agreement of the CPA and the disordered-lattice spectral functions indicates the tendency of delocalization.

In Fig.~\ref{fig:2}(b), we vary the disorder strength $W$ at a fixed $E=0.03$ where Bloch oscillation effect is weak. We observe that the width of $C(x)$ reaches a maximum at $W=0.4\sim 0.45$. This non-monotonic behavior is a result of an interplay of the number of accessible localized levels near the Fermi energy with increasing disorder~\cite{mozumdar2025}, and the localization of the electronic wave functions at very high disorder.

\begin{figure*}
\centering
\includegraphics[width=\textwidth]{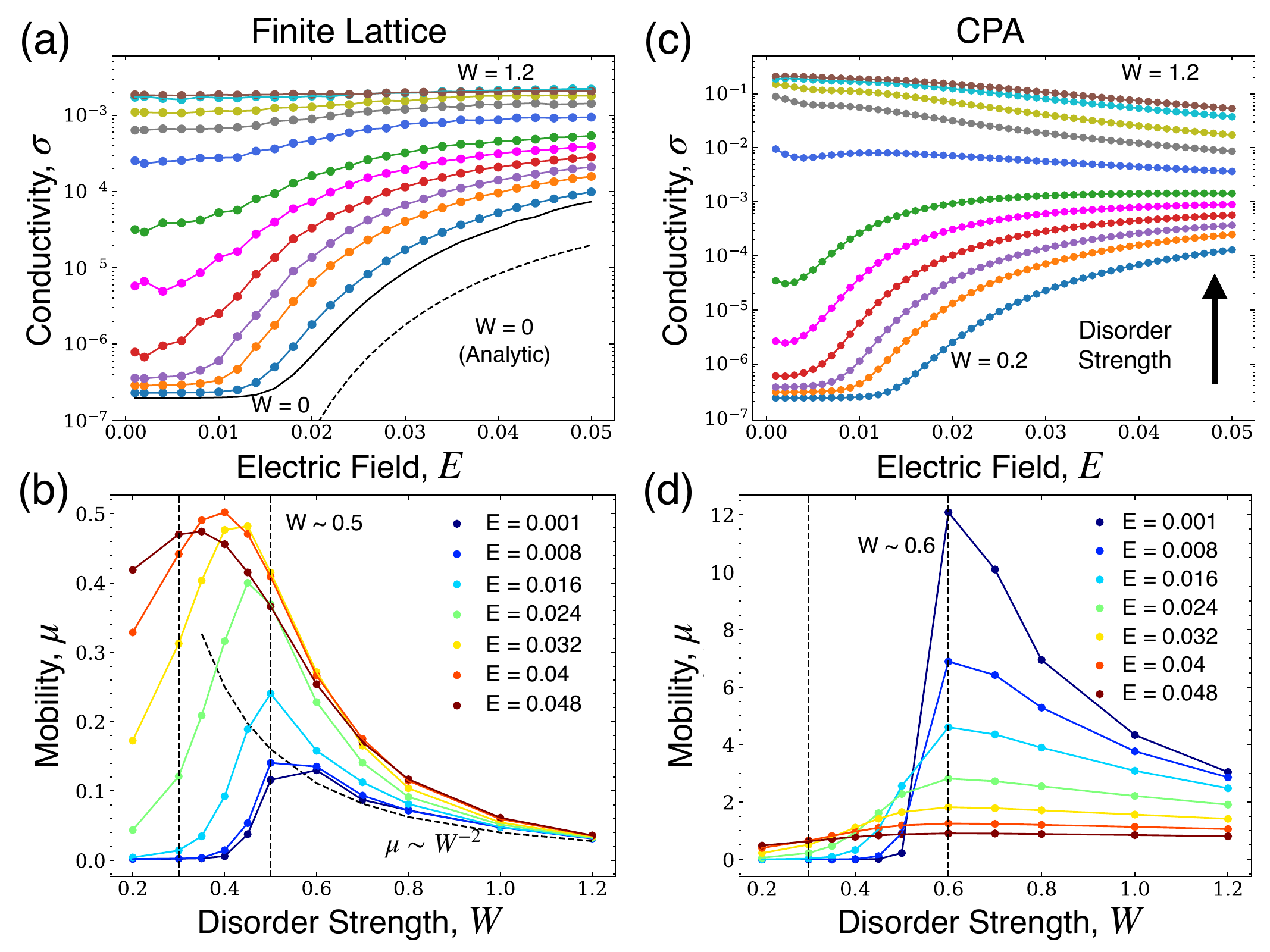}
\caption{(a) Conductivity vs electric field ($\Gamma=0.0005$, $N=501$, $t=1$, $T=0.01$). The average conductivity slowly increases with the electric field. In the low electric field region the conductivity strongly varies with the disorder strengths [$W \in \{0.2, 0.3, 0.35, 0.4, 0.45, 0.5, 0.6, 0.7, 0.8, 1.0, 1.2\}$: increasing in order indicated by the arrow in \ref{fig:3}(b)] with a rapid rise in the VRH regime $0.3 < W< 0.5$. The black solid and the dashed line show the numerical and analytic behaviors [see Eq. (\ref{app:J2})], respectively, at zero disorder.
(b) Mobility vs disorder strength ($\Gamma=0.0005$, $N=501$, $t=1$, $T_B=0.01$) for different electric fields. The average mobility at low electric fields $E < 0.04$  sharply rises in the disorder range $\Delta < W \lesssim 0.5$ marked by the black dashed lines for the VRH regime. Beyond this disorder range ($W > 0.5$) the mobility typically follows roughly the Fermi golden-rule ($\sim W^{-2}$) shown by the black dashed line. 
(c) CPA calculation of conductivity. While the overall behavior is consistent with the disordered-lattice calculation, the high-disorder limit strongly overestimates the conductivity.
(d) CPA calculation of mobility. Despite the similarity of the overall lineshape, the CPA overestimates the mobility and fails to produce the correct trends in the high-$W$ limit.
}
\label{fig:3}
\end{figure*}

The electron propagation under the electric field in Fig.~\ref{fig:2}(a-b) can be summarized by computing the drift distance $L_{\rm drift}$ defined as 
\begin{equation}
    L_{\rm drift} = \frac{\int_0^\infty C(x)xdx}{\int_0^\infty C(x)dx}.
\end{equation}
We make a note here that the drift distance is the travel distance of wave-propagation under the effect of an external field during the charge carrier's lifetime.  This is distinguished from the optimal hopping distance as proposed by Mott's formalism (discussed in later sections). Drift distance incidentally, could be a cumulative result of several hops over large distances in the lattice. As seen in Fig. \ref{fig:2}(c), the drift distance increases rapidly at small electric fields, and reaches a maximum as $E$ increases. Initially, $L_{\rm drift}(E)$ starts with a zero slope in the low disorder limit (blue, $W=0.2$) since localized states lie above the Fermi energy. In the range $0.02 < E < 0.04$ at $W=0.2$, the broadening of the band edge results in a rapid increase of the VRH transport. This is enabled by the tunneling of non-local reservoir electrons to localized states assisted by the electrostatic potential, similar to the Poole-Frenkel effect~\cite{FrenkelPR1938}. As the disorder increases, the threshold electric field required for the rapid growth of $L_{\rm drift}$ becomes smaller, reaching a critical condition at $W=0.4$ (red) - $0.45$ corresponding to the maximally VRH condition. As the disorder further increases ($W>0.5$), a significant fraction of sites are filled and the VRH mechanism gives way to strongly disordered transport with a much-reduced variation of $L_{\rm drift}$ by $E$, as discussed earlier. The crossover of the VRH transport to the Bloch oscillation behavior is most visible at the maximal VRH disorder at $W=0.4-0.45$, with the dashed line denoting the behavior of $x_B\propto 1/E$.

Fig.~\ref{fig:2}(d) shows the variation of $L_{\rm drift}$ with the disorder strength $W$. Within the VRH window (dashed vertical lines) of
\begin{equation} 
    \Delta < W < W^{*}\mbox{ with }W^*\approx 0.5
    \label{eq:vrh}
\end{equation}
the drift distance goes through a maximum at intermediate fields ($E\lesssim 0.04$). At high fields, the VRH behavior is ineffective and $L_{\rm drift}$ shows a monotonic behavior. The conclusion from the correlation study for the VRH range is fully consistent with that of the wave function via the inverse-participation-ratio~\cite{mozumdar2025}.

\subsection{\label{sec:level3-2}Nonlinear Conductivity in Disordered Lattice}

Now, we turn our attention to more directly measurable quantities of conductivity $\sigma(E)$ and mobility $\mu$.
In Fig.~\ref{fig:3}(a), we show the conductivity averaged over 2000 disorder configurations from the disordered-lattice calculations as a function of the electric field at different disorder strengths. At low disorder $W<0.4$ (red), the conductivity shows two distinct regimes, separated by the shoulders in the range $E=0.01\sim 0.02$. In the high-field regime, the conductivity shows the activation behavior by the electric field, mainly given by the Landau-Zener tunneling~\cite{hanPRB2018}. The exponential behavior of the conductivity is overridden by the coupling to the reservoirs, and it reaches a plateau as $E\to 0$. As discussed in previous work~\cite{mozumdar2025}, the conduction electrons hybridize with the reservoir continuum for the electron density $n\propto\Gamma$ with the acceleration time proportional to $\Gamma/\Delta$, leading to $\sigma(E\to 0)\propto\Gamma^2$ [Eq.~(19) of Ref.\cite{mozumdar2025}]. Despite the particular form of dissipation in our model, the coupling to the environment is universal, and the lower-bound behavior of the field-activation is generally expected.

As detailed in Appendix-\ref{App:1}, an analytic description (dashed line) of the activation behavior, Eq.~(\ref{app:J2}), shows a reasonable agreement apart from an overall factor. As the disorder increases, the region of the linear conductivity shrinks until we get into the VRH regime close to $W=0.4\sim \Delta$. Once $W\gtrsim \Delta$, the conductivity increases very rapidly over many orders of magnitude up to $W\lesssim 2\Delta$, agreeing with the VRH range, Eq.~(\ref{eq:vrh}). As the disorder increases further outside the VRH regime, the conductivity remains nearly independent of the field $E$, consistent with the Drude limit with a metallic weight as the system becomes delocalized.

We gain a deeper understanding by studying the mobility $\mu$, i.e. the average conductivity per carrier density ($\mu = \sigma/\langle n_{l} \rangle$ as defined in section-\ref{sec:level2-1}). The mobility, as shown in Fig.~\ref{fig:3}(b), maximizes in the VRH regime, defined in Eq.~(\ref{eq:vrh}), in direct correspondence with the drift distance $L_{\rm drift}$ [see Fig.~\ref{fig:2}(d)]. As argued earlier, the enhanced mobility results from the transport through disorder-generated ingap states before the disorder potential becomes detrimental to the transport. The main difference of the mobility behavior from that of the drift distance [see Fig.~2(d)] is that the mobility decreases as $W\to 0$. This can be understood as follows. While the travel distance by the Bloch oscillation contributes to $L_{\rm drift}$, such oscillations cancel out in the DC transport for the conductivity and mobility. Therefore, mobility gives more relevant information for disordered transport. The high disorder limit $\mu\sim W^{-2}$ (dashed line) is consistent with the Fermi-golden-rule interpretation.

\begin{figure*}
\centering
\includegraphics[width=\textwidth]{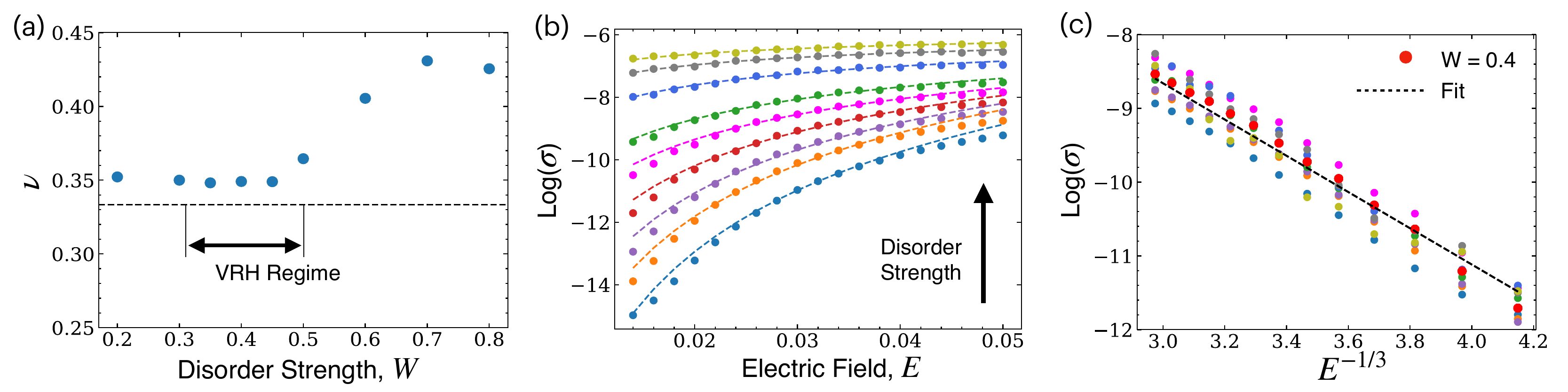}
\caption{(a) The exponent $\nu$ is evaluated using nonlinear curve-fitting as described in the text. The computed exponent is very close to the theoretical value of $\nu = 1/3$ depicted by the dashed black line for $W \lesssim W^{*}$. $\nu > 1/3$ for $W > W^{*}$ since the disorder values fall outside of the VRH Regime. (b) Log of Conductivity is plotted against the electric field for different disorder strengths [$W \in \{0.2, 0.3, 0.35, 0.4, 0.45, 0.5, 0.6, 0.7, 0.8\}$: increasing in the order indicated by the arrow] with a fit (dashed line) of Eq. (\ref{eq:nu13}). The fit exponent is computed as $\nu$. (c) $\text{Log}(\sigma)$ plotted against $E^{-1/3}$ which depicts a straight line for all disorder values. We scale all the curves to match the fit of $W =0.4$ to show a general agreement between the scaling law and the numerical result.}
\label{fig:4}
\end{figure*} 

We contrast the disordered-lattice calculations with the CPA method.  The results for conductivity in the CPA calculation are shown in Fig. \ref{fig:3}(c). For weaker disorder strengths ($W < 2\Delta = 0.6$), the behavior is similar to the lattice calculations, where the conductivity rises slowly at some value with respect to the electric field. For stronger disorder ($W > 2\Delta$), the CPA-conductivity is strongly overestimated compared to the disordered-lattice calculation, especially in the low-field limit. The insulator-to-metal transition occurs at $W = 0.6$ for our chosen values of $\Delta$ and $\Gamma$, with the CPA conductivity decreasing with the field in the high disorder limit, in contrast to (a). As shown in Fig. \ref{fig:3}(d), the mobility undergoes a sharp transition at $W = 2\Delta$. This is the disorder strength at which the CPA spectral function band edge crosses the Fermi level\cite{mozumdar2025}. In contrast, the mobility from the disordered-lattice calculations shows a gradual rise before capping at the critical disorder, where the Lifshitz tail smears into the Fermi sea at the zero-field limit. 

This contrast between the two calculations indicates the difference in capturing the physics of the model. While the CPA method is generally successful in capturing the spectral properties at very high fields, it is unable to capture the transport physics of Anderson localization at lower electric fields hence, it does not show any signatures of VRH, which typically arise from the transport in localized levels.

\subsection{\label{sec:level3-3}Electric-Field Scaling Law of VRH}

In this section, we discuss an extension of Mott's VRH behavior of Eq. (\ref{mott eq}) into the strong-field limit. The discussions on the strong-field scaling have been quite controversial. Differences in the $E$-field strength versus the temperature and the statistical criteria for hops under a field have led to competing predictions for the scaling exponent~\cite{Mott01071970,shklovskii1973,PhysRevLett.95.166604,Apsley1974-APSTFO}. Here, we present our own interpretation of the strong-$E$-field scaling and provide microscopic and unbiased lattice calculations to support our relation. The electric field accelerates the electrons, which can be excited to non-local sites in a chain. We modify Mott's model by incorporating the field-driven excitations in the Miller-Abraham's hopping probability ${\cal W}$, Eq.~(\ref{eq:MillerAbraham}), as 
\begin{equation}
    \mathcal{W} =\mathcal{W}_0 \exp\left[-\frac{2x}{\xi} - \frac{\epsilon}{eEx}\right],
    \label{eq:es1}
\end{equation}
with the hopping distance $x$ between two localized states of level-spacing $\epsilon$, the electric-field $E$, and the localization length $\xi$. The first term in the exponent captures the spatial overlap between two localized levels, and the second term is modified to incorporate electric field excitations in place of the thermal excitation $k_{\text{B}}T$, as demonstrated in a dissipative field-driven lattice model~\cite{hanPRB2013}. At low temperatures, the electric field is the major source of excitations in the system, and the scaling behavior may be dictated by the nonlinear electric field. Here, we imply that an electron gains the energy $eEx$ for a most probable hopping distance $x$ in the direction of the field, and after making the hop, the excess energy relaxes into thermal energy. The driven chain thermalizes by emission of hot-electrons followed by absorption of cold-electrons from the reservoirs, and by creating energetic electron-hole pairs in the infinite sea of electrons in the Fermion bath. It has been shown analytically~\cite{hanPRB2018} that the electron's effective temperature is linearly proportional to the field in the clean limit in the particle-hole symmetric metals and insulators. Our single-band insulator model, which lacks the particle-hole symmetry, shows a similar trend of increasing electron temperature, as observed from the smearing of the nonequilibrium distribution with increasing electric field in our previous study [Fig. 5(a) of \cite{mozumdar2025}].

In the steady-state limit, repeated inelastic scattering establishes an effective temperature as a function of the most probable hopping distance $x$. With this thermal description of the field excitation, we proceed analogously with Mott's mechanism of VRH, for the $E$-field scaling relation. We minimize the function in the exponent of the Miller-Abraham's relation, Eq.~(\ref{eq:es1})
\begin{equation}
    f(x,\epsilon) = \frac{2x}{\xi} + \frac{\epsilon}{eEx}
    \label{eq:es2}
\end{equation}
for some optimum values of $x$ and $\epsilon$ which maximize the overall hopping probability. Following Mott's statistical argument, if there is at least one probable hop in some spatial distance and energy window $\epsilon$, we can write in one-dimension
\begin{equation}
    \epsilon = \frac{1}{g(E_F)\cdot 2x}
    \label{eq:es3}
\end{equation}
where $g(E_F)$ is the density of states at the Fermi level $E_F$, and for simplicity we set $g(E_F)=1/(2Wa_0)$ with the unit-cell normalization with the lattice constant $a_0$. The exponent becomes
\begin{equation}
    f(x)=\frac{2x}{\xi}+\frac{Wa_0}{eEx^2}.
    \label{eq:fx}
\end{equation}
We can then optimize Eq. (\ref{eq:es2}) with respect to $x$ following similar steps to Mott's approach~\cite{mott1968conduction}, and obtain the optimal $x^*=(Wa_0\xi/eE)^{1/3}$, from which we obtain
\begin{equation} \label{eq:nu13}
\sigma(E)\sim\exp\left[-3\left(\frac{Wa_0}{eE\xi^2}\right)^{1/3}\right],
\end{equation}
with the exponent $\nu=1/3$. We may define the effective temperature $T_{\rm eff}$ at the most probable $x^*$ as $f(x^*)=eEx^*/T_{\rm eff}$ and obtain~\cite{marianer1992}
\begin{equation} 
    T_{\rm eff}(E)=\frac{1}{3}eE\xi.
\label{eq:teff}
\end{equation}
Discussions so far apply in the intermediate field regime where the balance between the two terms in Eq.~(\ref{eq:fx}) leads the probable hopping range $x^*$ to be comparable to $\xi$, or equivalently, $T_{\rm eff}=eE\xi/3\sim Wa_0/(3\xi)$ where the electron excitation energy by the field remains a fraction of the disorder potential strength $W$.

In a stronger field limit $E\to \infty$, however, the field becomes so strong that the potential drop by the field immediately makes most hops energetically favorable. This regime is reached when $Wa_0/(eEx^2)\sim 1$ or $x\sim\sqrt{Wa_0/eE}\ll\xi$, beyond which the thermal factor becomes irrelevant. This threshold $x$ determines the most likely hopping distance, and the Miller-Abraham's formula takes the first term of the wave-function overlap~\cite{Mott01071970,shklovskii1973}, and we approximate the conductivity as
\begin{equation}
    \sigma(E)\sim e^{-2x/\xi}=\exp\left[-2\sqrt{\frac{Wa_0}{eE\xi^2}}\right],
    \label{eq:nu12}
\end{equation}
with the exponent $\nu=1/2$, in agreement with Ref.~\cite{shklovskii1973} in the one-dimensional limit.

Now, we present the numerical results. As in previous calculations, we perform straightforward numerical calculations on a finite-size electronic lattice with the electrostatic potential created by the field, by using the Keldysh Green's function method. We then confirm the above prediction by performing a nonlinear curve-fit on our numerical simulation of the conductivity shown earlier in Fig.~\ref{fig:3}(a) to this scaling law in Eqs.~(\ref{eq:nu13}) and (\ref{eq:nu12}). We compute the scaling exponent $\nu$ by treating $\sigma_{\infty}$, $E_0$ and $\nu$ as free parameters. The exponent $\nu$ is shown in Fig. \ref{fig:4}(a) for different disorder strengths. The estimated exponent is close to the theoretical value of $1/3$ predicted in Eq. (\ref{eq:nu13}), which is marked by the dashed line in Fig. \ref{fig:4}(a). However, at strong disorders ($W > W^{*}$) where stronger electric fields are required for significant variations of conductivity, the numerically obtained exponent $\nu\approx 1/2$ becomes more consistent with Eq.~(\ref{eq:nu12}).

The nonlinear fit is depicted in Fig.~\ref{fig:4}(b), which shows $\log({\sigma})$ against the electric field, with Eq.~(\ref{eq:nu13}) shown as dashed lines. In the low-field limit, [$E < 0.01$, this range is omitted in Fig.~\ref{fig:4}(b)], the dissipation effects dominate over the VRH transport, and the conductivity is constant with electric field. While in the high-field limit ($E > 0.4$), the fit overestimates the numerical conductivity as the system exhibits Bloch oscillations. Therefore, we note that Eq.~(\ref{eq:nu13}) is a good fit at intermediate electric fields $0.13\lesssim E \lesssim 0.4$ where the VRH transport is observed. In Fig.~\ref{fig:4}(c), $\log({\sigma})$ vs $E^{-1/3}$ shows a linear behavior which captures the electric field scaling law in Eq.~(\ref{eq:nu13}) reasonably well in the intermediate field regime.

\begin{figure}
\includegraphics[width=\linewidth]{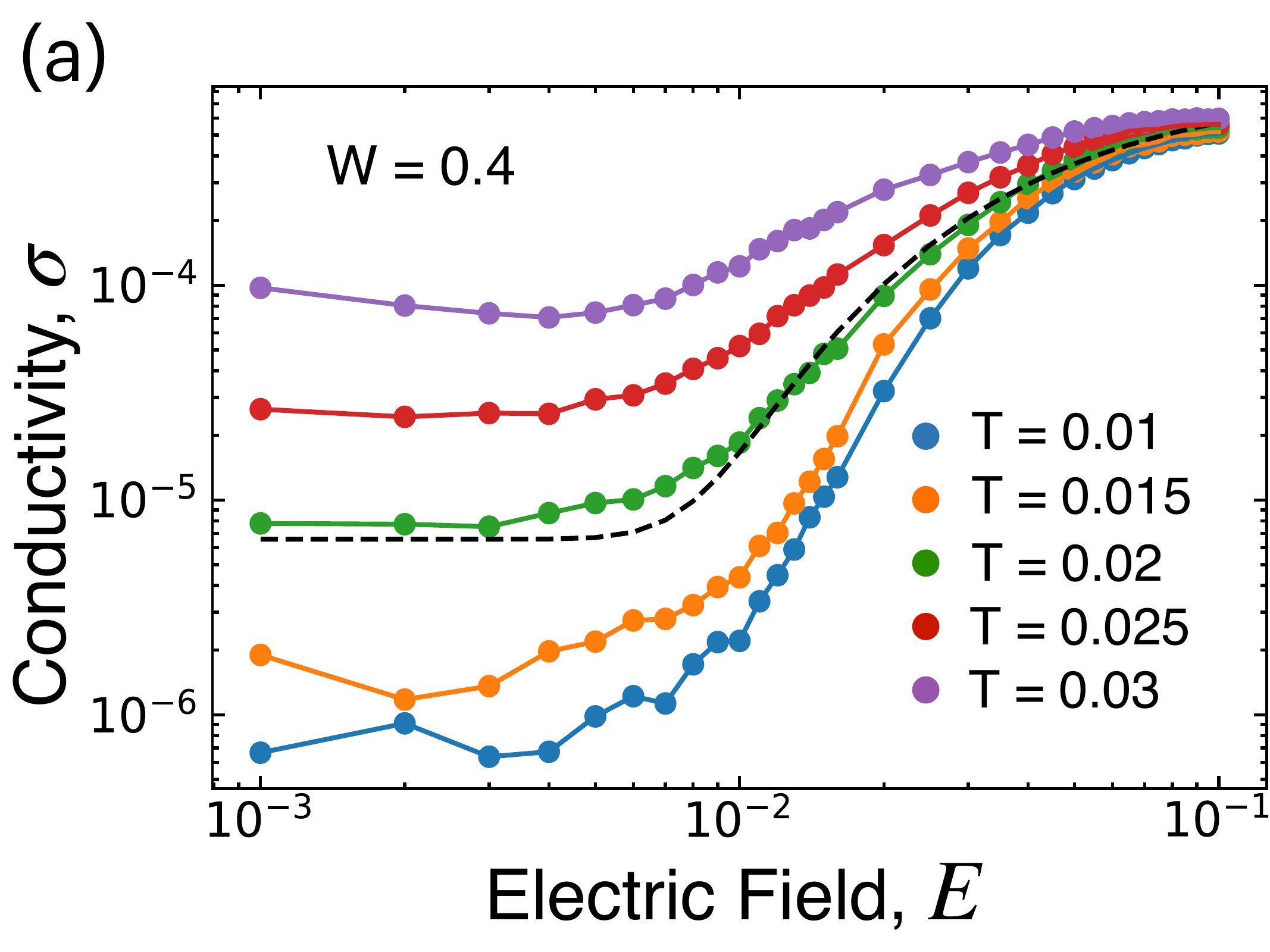}
\includegraphics[width=\linewidth]{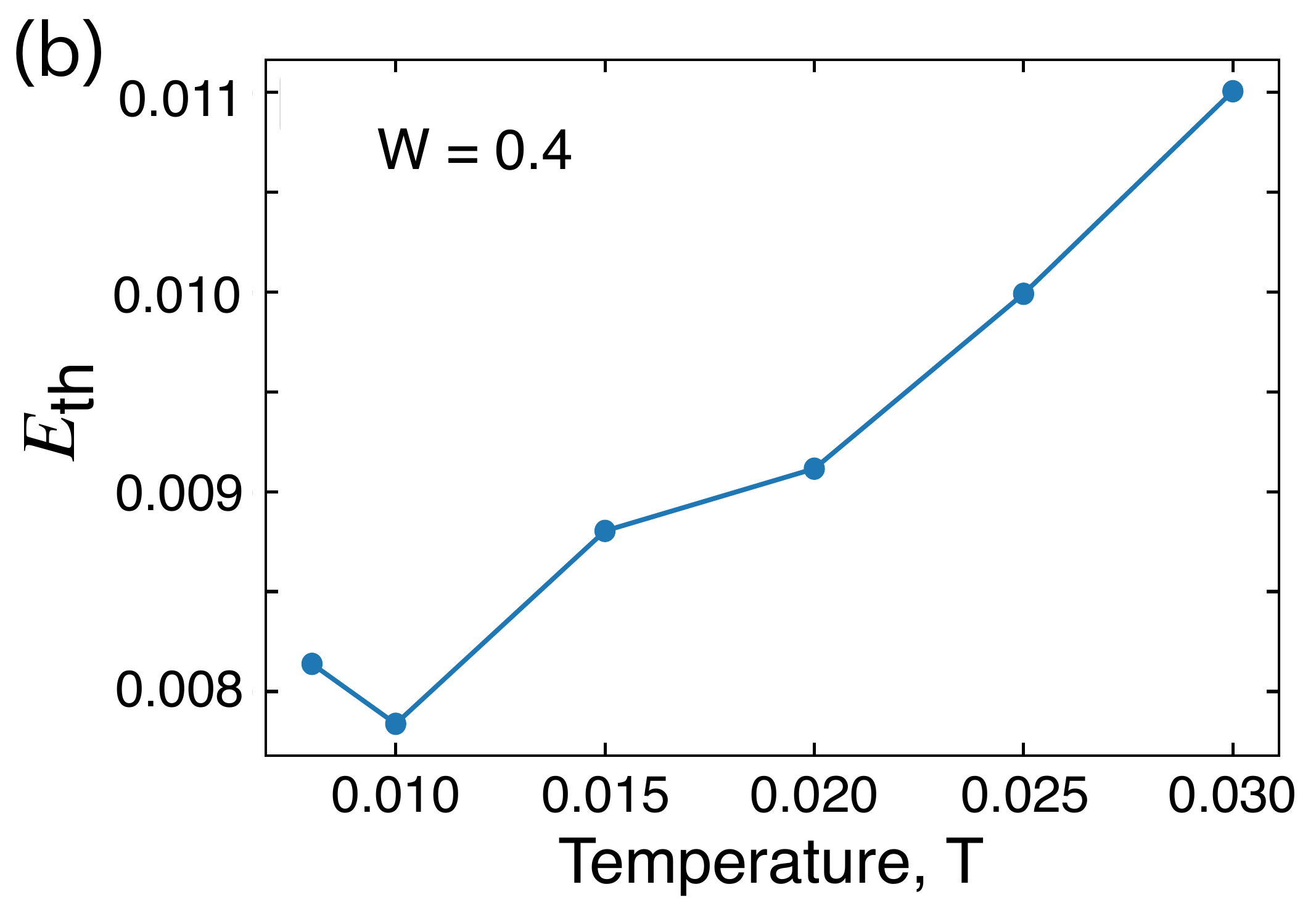}
\caption{(a) Conductivity vs electric field ($\Gamma=0.0005$, $N=501$, $t=1$, $W=0.4$) in the VRH regime for different Temperatures. In the low-field limit, the constant part of the conductivity follows Mott's $T^{-1/2}$-scaling of the VRH. Conductivity crosses over to a temperature-independent limit at high fields. The threshold field at which the conductance starts to deviate from the zero-field limit is evaluated using an empirical fit method detailed in the text, shown as the black dashed line for $T = 0.02$. The threshold field progresses to higher values as $T$ increases. (b) Threshold field $E_\text{th}$ vs temperature $T$. }
\label{fig:5}
\end{figure}  

Different experiments have reported different scaling exponents \cite{zaccone2016,PhysRevLett.92.216802,PhysRevB.40.3387,Rahman_2010,Park2017,PhysRevB.61.8108}, with the $E^{-1/3}$ scaling been observed in some experiments \cite{PhysRevB.52.5598, 10.1063/1.1421238} and $E^{-1/2}$ scaling most commonly reported~\cite{PhysRevB.40.3387, PhysRevLett.92.216802,Park2017}. Although various theoretical assumptions\cite{PhysRevLett.95.166604,zaccone2016,PhysRevB.70.235120,Apsley1974-APSTFO,rodin2009numerical} have been used to explain these scaling behaviors, based on material properties under study. However, the microscopic mechanism of variable range hopping under an electric field is still a matter of debate. Our calculations, apart from the choice of the model, are directly based on quantum mechanical equations of motion and offer insights into the nonlinear VRH transport in such low-dimensional disordered systems. The electronic model of a one-dimensional disordered chain presented in this study could be reflected in some low-dimensional disordered semiconductors~\cite{Randle2022}, which have been used to study nonlinear transport under a varied range of electric fields.

We conclude this section by presenting experimentally relevant temperature dependence of the VRH in the high-field regime. The temperature dependence of the threshold behavior is often considered an important clue to various resistive transitions for the debate of electronic versus thermal mechanism~\cite{janod2015,han2023correlated}. A log-log plot of the conductivity vs the electric field, shown in Fig.~5(a), displays two transport regimes separated by the crossover behavior. The low-field behavior of the constant conductivity is governed by the linear conductivity, which is consistent with Mott's $T^{-1/2}$-scaling. The high-field limit, however, is nearly independent of the temperature. This is again consistent with the view that a high $E$ destroys the localization. Constrained by the two limits, the temperature dependence of the localization-delocalization crossover is of strong experimental interest.

To evaluate the threshold electric field $E_{\rm th}$ as a function of temperature, we use an empirical fit motivated by the activation mechanism similar to the discussions in Appendix Eq.~(A6).
\begin{equation}
    \sigma(E,T) = \sigma_1(T) + \sigma_2\exp[-(E_c/E)]
\end{equation}
and we compute the $E_{\rm th}$ from the fit parameter $\sigma_1$, $\sigma_2$ and $E_c$ as 
\begin{equation}
    E_{\rm th} = \frac{E_c}{\ln[\sigma_2/\sigma_1]}.
\end{equation}
The $E_{\rm th}$ vs. $T$, as shown in Fig.~\ref{fig:5}(b), is an increasing relation. Such relations often suggest that a stronger electric field is required to overcome the thermal fluctuations, which means that the thermal effects counteract the nonequilibrium bias and it has been taken for evidence of non-thermal transitions~\cite{han2023correlated,Yuen2009,Rahman_2010} from many-body effects.

Our scenario, based on the disorder model, suggests something less dramatic. The $\sigma_1(T)$-term strongly varies with temperature over a few orders of magnitude due to the VRH mechanism, while the high-$E$ limit of the conductivity $\sigma(E,T)$ has much weaker $T$-dependence since the nonequilibrium-driven effective temperature would overwhelm the bath temperature. Then, by writing the conductivity as $\sigma_1(T)+\sigma'(E)$ where $\sigma'(E)$ is $T$-independent and an increasing function of $E$, the crossover would happen when $\sigma_1(T)\approx\sigma'(E_{\rm th})$, thus $E_{\rm th}$ is an increasing function of $T$. 

\section{\label{sec:level4}Conclusion}
We investigated the phenomenology of the nonlinear electron transport in a disordered insulator lattice in the steady-state limit under a large electric field. The Keldysh Green's function method was used to solve the steady-state nonequilibrium rigorously with the dissipation modeled by local fermion thermostats. We set up the lattice in the bulk limit with a large enough self-averaging disorder region. In the experimentally relevant energy scale with the electric field of $10-10^4$ kV/cm, we showed that the variable range hopping (VRH) is the dominant transport mechanism when the disorder parametrized by $W$ begins to overcome the gap $\Delta$ in the range $\Delta\lesssim W\lesssim 2\Delta$. 

In the VRH regime, the electron drift distance, electric conductivity, and mobility showed strong dependence on $E$ and $W$. In a narrow window of $E$, the hopping range varied more than an order of magnitude as a function of $E$ and $W$, with a maximum that divides the regime of the VRH and the Bloch oscillation. Conductivity and mobility demonstrated disorder-assisted transport with a huge enhancement over the VRH regime. A comprehensive comparison with the coherent-potential approximation (CPA) emphasized the role of non-local disorder physics in the disordered-lattice model.

The nonequilibrium extension of Mott's VRH scenario is confirmed by an electric-field scaling of the conductivity $\sigma(E)$ through an electronic model. For electric fields, outside the linear response and the Bloch oscillation regime, we verified the scaling law $\sigma(E)\sim \exp[-(E_0/E)^\nu]$ with $\nu=1/3$ and $\nu=1/2$ in the intermediate and strong disorder regime, respectively, in one-dimension. This can be attributed to electric field-driven electrons occupying and hopping over non-local localized sites in a similar energy range in the disordered active region. We envision that experimental verification of the nonlinear transport is one of the most pressing issues in nonequilibrium-disorder physics. While the VRH physics has been well documented in the linear regime, its extension to the strong field regime awaits further studies. The transport mechanism in transition-metal oxides, for instance, is often considered as mediated by transport through localized in-gap states, and the nonlinear $IV$ relation before the switching may be examined for the nonlinear transport scaling~\cite{zeng2006,PhysRevB.52.5598, PhysRevLett.92.216802,PhysRevB.40.3387,Rahman_2010}. Finally, we addressed the VRH mechanism for the unconventional temperature dependence of the conductance threshold, which can be experimentally examined in quasi-one-dimensional systems~\cite{Rahman_2010,Yuen2009}.

\begin{acknowledgments}
We thank J. P. Bird, J. Hofmann, G. Sambandamurthy, M. Randle, H. Zeng, and Z. Zhang for their helpful discussions. In particular we greatly benefited from insightful discussions with B. I. Shklovskii. We acknowledge computational support from the CCR at University at Buffalo. KM and HFF acknowledge the support from the Department of Energy, Office of Science, Basic Energy Sciences, under grant number DE-SC0024139. 
\end{acknowledgments}

\appendix

\section{ \label{App:1} Nonlinear Transport in Clean Limit}

Adopting a similar approach to our previous paper \cite{mozumdar2025}, we can write the local electron density $n_{\rm loc}$ in the electric field in terms of the same-time lesser Green's function, after the change of variable $p+ES\to p$ in the expression defined in Appendix A [Eq.(A6) of \cite{mozumdar2025}], as
\begin{eqnarray}
    n_{\rm loc}(E) &= &\frac{i\Gamma}{\pi}\int\frac{dp}{2\pi}\int_{-\infty}^\infty ds\int_{-\infty}^{-|s|/2}dS\frac{e^{2\Gamma S}}{s+i\eta}\times \nonumber \\
    &\times &\exp\left[is\epsilon(p)+\frac{iE^2s^3}{24m}\right] \\
    & = & \frac{i}{4\pi^2}\int dp\int_{-\infty}^\infty ds\frac{e^{-\Gamma |s|}}{s+i\eta}
    \exp\left[is\epsilon(p)+\frac{iE^2s^3}{24m}\right].
    \nonumber
\end{eqnarray}
The Gaussian integral over $p$ and an integral over $S$ results in the expression
\begin{equation}
    \frac{ie^{i\pi/4}m^{1/2}}{(2\pi)^{3/2}}\int_{-\infty}^\infty ds\frac{e^{-\Gamma|s|}}{(s+i\eta)^{3/2}}
    \exp\left[i\Delta s+\frac{iE^2s^3}{24m}\right].\nonumber
\end{equation}
Rescaling the variable $s\to(\sqrt{8m\Delta}/E)s$, we obtain
\begin{equation}
    ie^{i\pi/4}\left(\frac{\sqrt{m}E}{(2\pi)^3\sqrt{8\Delta}}\right)^{1/2}\int_{-\infty}^\infty ds\frac{e^{-\gamma|s|}}{(s+i\eta)^{3/2}}e^{i\lambda(s+s^3/3)},\nonumber
\end{equation}
with $\lambda=\sqrt{8m\Delta^3}/E$ and $\gamma=\Gamma\sqrt{8m\Delta}/E$. We are interested in the limit of $\lambda \gg 1$ and $\gamma\ll 1$. By using the steepest descent technique for the $\lambda \gg 1$ limit, the integral simplifies to
\begin{eqnarray}
n_{\rm loc}(E) & = & \left(\frac{\sqrt{m}E}{(2\pi)^3\sqrt{8\Delta}}\right)^{1/2}
\sqrt{\frac{\pi}{\lambda}}e^{-\frac23\lambda} \nonumber  \\
&=& \frac{E}{8\pi\Delta}\exp\left[-\frac{2\sqrt{8m\Delta^3}}{3E}\right].
\label{app:n}
\end{eqnarray}

For the electric current in high field, however, we have to be cautious with the approximation of using the quadratic dispersion relation instead of that from a lattice model. Unlike the calculation of the occupation number above, the mechanical momentum $p+eEt$ increases beyond the first Brillouin zone in solids, and the continuum approximation will deviate from the lattice results, which will be discussed below. The electric current density $J$ is calculated as
\begin{eqnarray}
    J(E) & = & \frac{iE}{8\pi^2m\Gamma}\int dp\int_{-\infty}^\infty ds\frac{(1+\Gamma|s|)e^{-\Gamma|s|}}{s+i\eta} \nonumber \\
    & & \times \exp\left[is\epsilon(p)+\frac{iE^2s^3}{24m}\right] \\
    & = & \frac{e^{3\pi i/4}}{8\pi m}\left(\frac{\sqrt{2m}E^3}{2\pi
    \sqrt{\Delta}\Gamma^2}\right)^{1/2} \nonumber \\
    & & \times\int_{-\infty}^\infty ds\frac{1+\gamma|s|}{(s+i\eta)^{3/2}}e^{-\gamma|s|+i\lambda(s+s^3/3)}.
    \nonumber
\end{eqnarray}
The integral can be approximated by the steepest descent method up to the leading orders of $\gamma$ as
\begin{eqnarray}
& & \int_{-\infty}^\infty ds\frac{1-\frac12\gamma^2s^2}{(s+i\eta)^{3/2}}e^{i\lambda(s+s^3/3)} \nonumber \\
& \approx & e^{-3i\pi/4}\left(1+\frac{\gamma^2}{2}\right)\sqrt{\frac{\pi}{\lambda}}e^{-\frac23\lambda},
\end{eqnarray}
and
\begin{equation}
    J(E) \approx\frac{E^2}{8\pi m\Delta\Gamma}\left(1+\frac{4m\Delta\Gamma^2}{E^2}\right)\exp\left[-\frac{2\sqrt{8m\Delta^3}}{3E}\right].
    \label{app:J}
\end{equation}
The first term in the parenthesis corresponds to the Drude current $J_D=n_{\rm loc}(E)E/(m\Gamma)$ in the continuum model with $n_{\rm loc}(E)$ from Eq.~(\ref{app:n}). However, in the nonlinear regime, a lattice model at $E\gg\Gamma$ develops the Bloch oscillation with the period $T=\pi/(eEa)$ (we use the unit $e=a=1$). With a scattering time much exceeding this time scale, the Bloch oscillation averages out the current. In the regime of our interest [see Fig.~3(a)], $E/\Gamma \gtrsim 20$ after nonlinear threshold fields, and the Bloch oscillation eliminates the first contribution in Eq.~(\ref{app:J}). Therefore, the conductance in the lattice model can be approximated with the next leading order of $\Gamma$ as
\begin{equation}
    \sigma(E) \approx\frac{\Gamma}{2\pi E}\exp\left[-\frac{2\sqrt{8m\Delta^3}}{3E}\right].
    \label{app:J2}
\end{equation}
The result can be expressed as
$\sigma(E)=4(\Delta\Gamma/E^2)n_{\rm loc}(E)$. In the metallic lattice~\cite{hanPRB2013}, the conductivity in the strong Bloch oscillation limit $(E\gg\Gamma)$ is rigorously given as $\sigma_{\rm metallic}(E)=(8/\pi)(t\Gamma/E^2)n_{\rm metallic}(E)$ in the half-filled limit $n_{\rm metallic}(E)=1/2$, following a form similar to the above.
While Eq.~(\ref{app:J2}) may be improved by considering the tight-binding dispersion relation for comparison with the numerical results, Fig.~\ref{fig:3}(a) shows that Eq.~(\ref{app:J2}) matches the numerical nonlinear conductivity reasonably well after the nonlinear threshold. The leading linear dependence in $\Gamma$ is consistent with numerical results.
By equating Eq.~(A12) of \cite{mozumdar2025} to Eq.~(\ref{app:J2}) we may estimate the nonlinear threshold field at which the linear behavior crosses over to the nonlinear regime from the relation $\Gamma/(2\Delta)\sim \lambda e^{-(2/3)\lambda}$.

\nocite{*}

\bibliography{vrh2}

\end{document}